\documentclass{article}
\usepackage[english]{babel}
\usepackage[utf8]{inputenc}
\usepackage{microtype}
\usepackage{lmodern}
\usepackage{hyperref}
\usepackage{amsmath}
\usepackage{amssymb}
\usepackage{multicol}
\usepackage{mathtools}
\usepackage[thref,thmmarks,amsmath,hyperref]{ntheorem}
\usepackage{bbold}
\usepackage{enumitem}

\usepackage{csquotes}
\usepackage[backend=biber,natbib=true,url=false]{biblatex}
\usepackage{tikz}
\usetikzlibrary{arrows.meta}

\DeclareMathOperator{\tr}{tr}

\newcommand{\IN}{\ensuremath{\mathbb{N}}}
\newcommand{\IZ}{\ensuremath{\mathbb{Z}}}
\newcommand{\IR}{\ensuremath{\mathbb{R}}}
\newcommand{\IC}{\ensuremath{\mathbb{C}}}

\newcommand{\Hamiltonian}{\mathbb{H}}
\newcommand{\card}[1]{{\ensuremath{\left|#1\right|}}}
\newcommand{\imag}{\mathrm{i}}
\newcommand{\CAR}[1]{{\ensuremath{\IC[c,c^*]}}}

\newcommand{\one}{\mathbb{1}}
\newcommand{\N}{\mathbf{n}}
\newcommand{\C}{\mathbf{c}}
\newcommand{\Cd}{\C^*}
\newcommand{\PowerSet}[1]{\mathfrak{P}\left(#1\right)}
\newcommand{\HS}{{\mathcal{L}^2(\FockSpace)}}
\newcommand{\HSbasis}{\mathfrak{B}}
\newcommand{\HSrealBasis}{\mathfrak{B}^\IR}
\newcommand{\HilbertSpace}{\ensuremath{\mathfrak{h}}}
\newcommand{\FockSpace}{\mathcal{F}}
\newcommand{\Vacuum}{\Omega_\FockSpace}
\newcommand{\sign}[2]{\begin{bmatrix}\begin{matrix}#1\end{matrix}\\\begin{matrix}#2\end{matrix}\end{bmatrix}}

\newcommand{\SP}[2]{{\langle{#1}\mid{#2}\rangle}}
\newcommand{\dotcup}{\mathrel{\dot{\cup}}}
\newcommand{\DensityMatrices}{\mathcal{P}_1}

\newcommand{\kbOp}[1][k]{{\ensuremath{\mathcal{O}_{#1}(\mathcal{F})}}}
\newcommand{\PkbOp}[1][k]{{\ensuremath{\pi_{#1}}}}
\newcommand{\kbOpBasis}[1][k]{\mathfrak{B}_{#1}}

\newcommand{\kbOb}[1][k]{{\ensuremath{\mathcal{O}_{#1}^\mathbb{R}(\mathcal{F})}}}

\author{Bach, Volker \and Rauch, Robert}
\title{Orthogonalization of fermion $k$-Body operators and representability}

\begin{document}
\maketitle

\begin{abstract}
The reduced $k$-particle density matrix of a density matrix on finite-dimensional, fermion Fock
space can be defined as the image under the orthogonal projection in the Hilbert-Schmidt geometry
onto the space of $k$-body observables. A proper understanding of this projection is therefore
intimately related to the \emph{representability problem}, a long-standing open problem in
computational quantum chemistry. Given an orthonormal basis in the finite-dimensional one-particle
Hilbert space, we explicitly construct an orthonormal basis of the space of Fock space operators
which restricts to an orthonormal basis of the space of $k$-body operators for all $k$.
\end{abstract}

\maketitle

\section{Introduction}
\subsection{Motivation: Representability problems}
In quantum chemistry, molecules are usually modeled as non-relativistic many-fermion systems
(Born-Oppenheimer approximation). More specifically, the Hilbert space of these systems is given by
the fermion Fock space $\FockSpace=\FockSpace_f(\HilbertSpace)$, where $\HilbertSpace$ is the
(complex) Hilbert space of a single electron (e.g. $\HilbertSpace=L^2(\IR^3)\otimes \IC^2)$, and the
Hamiltonian $\Hamiltonian$ is usually a two-body operator or, more generally, a $k$-body operator on
$\FockSpace$. A key physical quantity whose computation is an important task is the ground state
energy
\begin{equation}
    \label{2def9341}
    E_0(\Hamiltonian)\doteq\inf_{\varphi\in\mathcal{S}}\varphi(\Hamiltonian)
\end{equation}
of the system, where $\mathcal{S}\subseteq\mathcal{B}(\FockSpace)'$ is a suitable set of states on
$\mathcal{B}(\FockSpace)$, where $\mathcal{B}(\FockSpace)$ is the Banach space of bounded operators
on $\FockSpace$ and $\mathcal{B}(\FockSpace)'$ its dual. A direct evaluation of \eqref{2def9341} is,
however, practically impossible due to the vast size of the state space $\mathcal{S}$.

\paragraph{Abstract representability problem}
As has been widely
observed, this problem can be reduced drastically by replacing the states
$\tau\in\mathcal{S}$ by a quantity $r_\tau$, the \emph{$k$-body reduction} of $\tau$, that
only encodes the expectation values of $k$-body operators in the state $\tau$. More precisely,
denote by $\kbOp\subseteq\mathcal{B}(\FockSpace)$ the subspace of $k$-body operators on $\FockSpace$
and let $\tau\in\mathcal{B}(\FockSpace)'$, then $r_\tau$ can be defined as the
restriction $\tau|_{\kbOp}\in\kbOp'$. In other words, if $i_k:\kbOp\to\mathcal{B}(\mathcal{F})$
denotes the inclusion map then the mapping $\tau\mapsto r_\tau$ is given by the dual map
$i_k':\mathcal{B}(\mathcal{F})'\to\kbOp'$, which we call the \emph{$k$-body reduction map}. Now, if
$\Hamiltonian\in\kbOp$ then $\tau(\Hamiltonian)=(i_k'\tau)(\Hamiltonian)$ for all
$\tau\in\mathcal{B}(\FockSpace)'$ and \eqref{2def9341} can be rewritten as
\begin{equation}
    \label{4dac81aa}
    E_0(\Hamiltonian)=\inf_{\tau\in\mathcal{S}}\tau(\Hamiltonian)
    =\inf_{\tau\in\mathcal{S}}r_\tau(\Hamiltonian)
    =\inf_{r\in i_k'(\mathcal{S})}r(\Hamiltonian),
\end{equation}
thus the evaluation of \eqref{2def9341} is, in principle, simplified, because the infimum has to be
taken over the much smaller set $i_k'(\mathcal{S})$. To explicitly compute the
right hand side of \eqref{4dac81aa} however, one has to find an efficient parametrization of the set
$i_k'(\mathcal{S})$. The \emph{representability problem for $\mathcal{S}$} (and $k\in\IN_0$)
amounts to characterize the image $i_k'(\mathcal{S})$ of \emph{representable} functionals on $\kbOp$
in a computationally efficient way.

\paragraph{Traditional representability problems}
The general framework of representability problems as discussed here is usually invisible in the
pertinent literature, because in concrete applications $\mathcal{S}$ is almost always chosen to be
(a subset of) the set of density matrices on $\FockSpace$ and $\kbOp'$ is identified with a suitable
subspace of $\mathcal{B}(\FockSpace)$. Moreover, in applications of physics or chemistry the by far
most important case is $k=2$, as the Hamiltonian usually is a two-body operator. In this case the
two-body reduction $i_k'(\rho)$ of an $N$-particle density matrix can be identified with the
(customary) $2$-RDM, which is a bounded operator on $\bigwedge^2\HilbertSpace$.

\paragraph{Erdahl's representability framework}
In this paper, only the case $\dim\HilbertSpace<\infty$ is considered, which is sufficient for many
important applications. For example, in quantum chemistry one commonly starts by choosing a finite
subset of $L^2(\IR^3)\otimes\IC^2$ of \emph{spin orbitals} and then considers their span
$\HilbertSpace$. In the finite-dimensional case, the reduced $k$-body reduction of a density matrix
$\rho$ can be introduced as the image $\PkbOp(\rho)$ under
the orthogonal projection onto $\kbOp$ \citep[see][]{Erdahl1978},
\begin{equation}
    \label{50476cd0}
    \PkbOp:\HS\to\kbOp\subseteq\HS.
\end{equation}
As it turns out, in the finite-dimensional case $\PkbOp$ is an equivalent description of the map
$i_k'$ introduced above. The reason for this is that in the finite-dimensional case
$\mathcal{B}(\FockSpace)=\HS$, where $\HS$ denotes the Hilbert space of Hilbert-Schmidt operators on
$\FockSpace$, and we may identify $\mathcal{B}(\FockSpace)'\cong\HS$ and
$\kbOp'\cong\kbOp$ via the Riesz isomorphisms. Under these identifications, the $k$-body reduction
map $i_k'$ is given by the adjoint $i_k^*$ of $i_k$ and $\pi_k=i_ki_k^*$. This geometric
interpretation of the representability problem is visualized in Fig. \ref{b61f9199}. Note that
Erdahl's representability framework breaks down in the infinite-dimensional case, because then
$k$-body operators are generally not Hilbert-Schmidt anymore.

\subsection{Related work}\label{50477cd0}
The idea of replacing density matrices by their reduced density matrices to simplify the
evaluation of \eqref{2def9341} can be traced back to Husimi \cite{Husimi1940}. First extensive
analyses were carried out in the 1950's and 1960's and lead, e.~g., to the solution of the
representability problem for one-body reduced density matrices of $N$-particle density matrices
\cite{Coleman1963,Garrod1964,Yang1962} and the development of (still very inaccurate) lower bound methods based on
representability conditions. In 1978 Erdahl introduced a new class of representability conditions
\cite{Erdahl1978}, which were found to significantly increase the accuracy of lower bound methods
\cite{Cances2006}. In 2005 the representability problem for the one-body reduced density matrices of
\emph{pure} states was solved by Klyachko \cite{Klyachko2006} based on results from quantum
information theory. In 2012 Mazziotti established a hierarchy of representability conditions
providing a formal solution of the representability problem for the two-body RDMs of $N$-particle
density matrices \cite{Mazziotti2012}. However, the general representability problem has been found
to be computationally intractible \cite{Mazziotti2012}, even on a quantum computer \cite{Liu2006}.
Computational advances \cite{Mazziotti2011} enabled a range of recent applications
\cite{Safaei2018,Sajjan2018,Montgomery2018}. Representability methods have also proved useful in
Hartree-Fock theory \cite{Bach2012}. For a more detailed overview on the history of representability
problems, we refer to \cite{Mazziotti2007} and \cite{Coleman2000}.

\subsection{Goal and main results}\label{abfc23c5}
\begin{figure}
    \centering
    \begin{tikzpicture}[scale=.7]
        \filldraw[fill=gray!20,very thick] (0,0) -- (2,0.5) -- (3,2.5) .. controls (3.2,4.5) .. (1,5)
        -- (-1,4) -- (-1.75,2) -- cycle;
        \draw (.8,2.5) node {$\DensityMatrices$};

        \draw[-Stealth] (4,2.5) -- node[above] {$\PkbOp$}(7,2.5);

        \draw (8,-0.5) node[below] {$\kbOp$} -- (8,5.5);

        \draw[dashed,very thin] (1,5) -- (8,5);
        \draw[dashed] (0,0) -- (8,0);

        \draw[Stealth-Stealth, very thick] (8,2) -- node[left, near end]{$\HSbasis$} (8,0) -- (6,0);
    \end{tikzpicture}
    \caption{Geometric interpretation of the representability problem for density matrices in finite
    dimensions: the mapping of density matrices $\rho\in\DensityMatrices$ to its $k$-body reduction
    as orthogonal projection $\PkbOp$ onto the subspace $\kbOp\subseteq\HS$ of $k$-body operators.
    The representability problem amounts to find an efficient characterization of the image
    $\PkbOp(\DensityMatrices)$ within $\kbOp$. The orthonormal basis $\HSbasis$ given in Theorem
    \ref{f5744217} is adapted to this situation as it restricts to an orthonormal basis
    $\HSbasis\cap\kbOp$ of $\kbOp$ for every $k\in\IN_0$.}
    \label{b61f9199}
\end{figure}
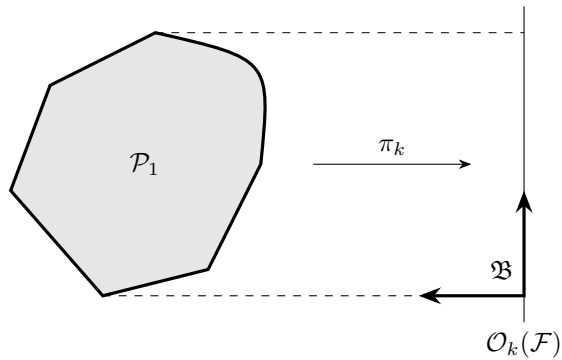

The goal of the present work is to shed more light on the projection $\PkbOp$ in the
finite-dimensional case. As a result, we explicitly diagonalize the orthogonal projections $\PkbOp$
simultaneously for all $k\in\IN_0$. More specifically, we prove the
following.\footnote{See Fig. \ref{b61f9199} for a geometric interpretation of
this result and its relation to the representability problem.}
\begin{theorem}[Main Theorem]\label{f5744217}
    Let $\dim_\IC\HilbertSpace=n<\infty$ and $\varphi_1,\ldots,\varphi_n$ be an orthonormal basis of
    $\HilbertSpace$. For $I=\{i_1<\ldots<i_j\}\subseteq\{1,\ldots, n\}$ define $\C_I\doteq
    c(\varphi_{i_j})\cdots c(\varphi_{i_1})$ and $n_I\doteq\Cd_I\C_I$, where $c(\varphi)$ denotes
    the usual fermion annihilation operator. Then the following is found
    \begin{enumerate}
        \item An orthonormal basis $\HSbasis$ of $\HS$ is given by the elements
        \begin{equation}
            \label{abfd23c5}
            \frac{1}{\sqrt{2^{n-\card{I\dot\cup J}}}}\sum_{A\subseteq L}(-2)^{\card{A}}\N_A\Cd_I\C_J,
        \end{equation}
        where $I,J,L$ run over all mutually disjoint subsets of $\{1,\ldots,n\}$.
        \item For any $k\in\IN_0$, $\HSbasis\cap\kbOp$ is an orthonormal basis of $\kbOp$.
    \end{enumerate}
\end{theorem}

Orthogonal decompositions of $\HS$ as implied by \thref{f5744217} have already been introduced,
e.~g., in \cite[Sec. 8]{Erdahl1978}, where an orthogonal decomposition
$\mathcal{B}(\mathcal{F})=\bigoplus_{n,m}\Lambda(n,m)$ is used to derive new classes of
representability conditions. The spaces $\Lambda(n,m)$ are generated by elements of the form
\eqref{b03562f0}, see Sec. \ref{a605ad65}. The orthonormal basis elements given in
\thref{f5744217}, however, have the additional property of being \emph{normal ordered}, which can be
used to express $\pi_k(\rho)$ in terms of the customary reduced density
matrices, as in the following example.

\begin{corollary}\label{f5744223}
Let $\rho$ be a particle number-preserving density matrix, $\gamma\in\mathcal{B}(\HilbertSpace)$ its $1$-RDM and $d\Gamma(\gamma)=\sum_{i,j}\gamma_{ji}c_i^*c_j$ the (differential) second quantization of $\gamma$. Then
\begin{equation}
	\label{f5744218}
	2^n\pi_1(\rho)=(n+1)-2\tr\{\gamma\}-2\hat{\IN}+4d\Gamma(\gamma),
\end{equation}
where $\hat{\IN}=\sum_ic_i^*c_i$ denotes the particle number operator.
\end{corollary}
A similar formula for $\pi_2(\rho)$ exists, but is much more complicated.

\subsection{Overview of the paper}

In Sec.~\ref{a0929406}, we introduce the necessary terminology and notation of
fermion many-particle systems and general density matrix theory, as well as,
some features specific to the finite-dimensional setting. In
Sec.~\ref{9861ec75}, we compute the Hilbert-Schmidt scalar product of
specific monomials in creation and annihilation operators (\thref{0aad}). In
Sec.~\ref{280ac894} we prove \thref{f5744217} in two steps, as follows.
\begin{enumerate}
    \item The orthonormal basis $\HSbasis$ of $\HS$ is constructed in \thref{9faf7dc5}.
    \item In \thref{46f8bbb2} we show that $\HSbasis\cap\kbOp$ is a basis of $\kbOp$ for all
    $k\in\IN_0$.
\end{enumerate}
In many cases one also considers the space $\kbOb$ of \emph{selfadjoint} $k$-body operators. We
generalize the above results in \thref{55efa9f6}, where we apply a suitable unitary transformation
$U$ on $\HS$ and show that the orthonormal basis $U(\HSbasis)$ of $\HS$ restricts to an orthonormal
basis of $\kbOb$ for all $k\in\IN_0$. Finally, in Sec.~\ref{a605ad65} we
present an alternative approach for constructing an orthonormal basis of $\HS$
with properties as in \thref{f5744217}, which was first communicated to us by
Gosset\footnote{\href{mailto:dgosset@uwaterloo.ca}{dgosset@uwaterloo.ca}} and
turned out to be already present in \cite{Erdahl1978}.

\subsection{Motivating application}

We illustrate the virtue of having orthonormal bases of the space of operators explicitly available
on the following example: Consider a fermionic many-particle system with finite-dimensional
one-particle Hilbert space $\mathfrak{h}$, a two-body Hamiltonian of the form
\begin{equation}
    \label{46575df0}
    \Hamiltonian=\sum_{i,j}t_{ij}c_i^*c_j+\frac{1}{2}\sum_{i,j,k,l}V_{ij;kl}c_i^*c_j^*c_lc_k,
\end{equation}
where $V_{ij;kl}\doteq\SP{\varphi_i\otimes\varphi_j}{V(\varphi_k\otimes\varphi_l)}$ is a matrix
element of a repulsive two-body potential $V\ge 0$. Let $\mathcal{B}$ be an orthonormal basis of
$\HS$. Then for any $\mathcal{A}\subseteq\mathcal{B}$ we have
$P_{\mathcal{A}}\doteq\sum_{\theta\in\mathcal{A}}\left|\theta\rangle\langle\theta\right|\le
\sum_{\theta\in\mathcal{B}}\left|\theta\rangle\langle\theta\right|=\mathbb{1}_\HS$ and, under
suitable positivity requirements on the potential $V$, we obtain
\begin{equation}
    \Hamiltonian\ge\sum_{i,j}t_{ij}c_i^*c_j+
    \frac{1}{2}\sum_{i,j,k,l}V_{ij;kl}c_i^*c_j^*P_\mathcal{A}c_lc_k
    \doteq\Hamiltonian_\mathcal{A}.
\end{equation}
Thus $E_0(\Hamiltonian_\mathcal{A})$ is a lower bound, which are usually more difficult to derive
than upper bounds, for the ground-state energy $E_0(\Hamiltonian)$ of the original quantum system.
In many situations, after a suitable choice of an orbital basis $\varphi_1,\ldots,\varphi_n$ of
$\HilbertSpace$, the orthonormal basis $\HSbasis$ given by \thref{f5744217} and a suitable choice of
$\mathcal{A}\subset\mathcal{B}$ leads to a nontrivial lower bound $E_0(\Hamiltonian_\mathcal{A})$ of
$E_0(\Hamiltonian)$.

\section{Foundations}\label{a0929406}
Throughout this work, $\mathfrak{h}$ denotes the one-particle Hilbert space, i.e., a separable complex
Hilbert space. We consider only the \emph{finite-dimensional case} here and assume
$n\doteq\dim_\IC\mathfrak{h}<\infty$ throughout the paper.

\subsection{General notions}
In this subsection, we will recall some relevant notions from general density matrix theory of
fermion many-particle systems that are also valid when $\dim\mathfrak{h}=\infty$.

\paragraph{Hilbert spaces}
If not stated otherwise, all Hilbert spaces are assumed to be complex. For a Hilbert space
$\mathcal{H}$, the inner product between elements $\varphi,\psi\in\mathcal{H}$ is denoted by
$\SP{\varphi}{\psi}_\mathcal{H}$ and is assumed to be \emph{anti-linear} in the first and
\emph{linear} in the second component. When there is no risk of confusion, we will freely omit the
subscript $\mathcal{H}$ of the inner product. By $\mathcal{B}(\mathcal{H})$ we denote the C*-algebra
of linear bounded operators on $\mathcal{H}$.

\paragraph{Hilbert-Schmidt operators}
The space of Hilbert-Schmidt operators on a Hilbert space $\mathcal{H}$ is denoted by
$\mathcal{L}^2(\mathcal{H})$ and is a Hilbert space with respect to the inner product
$\SP{a}{b}_{\mathcal{L}^2(\mathcal{H})}\doteq\tr\{a^* b\}$. Furthermore, $\HS$ is endowed with a
natural real structure (i.e., a complex conjugate involution) given by the Hermitian adjoint.

\paragraph{Fermion Fock space}
For a Hilbert space $\mathfrak{h}$, the associated \emph{fermion Fock space}
$\mathcal{F}\doteq\mathcal{F}(\mathfrak{h})$ is the completion of the Grassmann algebra
$\bigwedge\mathfrak{h}=\bigoplus_{k\ge 0}\bigwedge^k\mathfrak{h}$ with respect to the inner product
defined by
\begin{equation}
    \label{823d830d}
    \SP{\varphi_1\wedge\cdots\wedge\varphi_k}{\psi_1\wedge\cdots\wedge\psi_l}\doteq
    \begin{cases}
        \det\left(\SP{\varphi_i}{\psi_j}\right)_{i,j=1}^k&\text{if }k=l,\\
        0&\text{otherwise.}
    \end{cases}
\end{equation}
The neutral element $1\in\IC\doteq\bigwedge^0\mathfrak{h}\subset\FockSpace$ of the wedge product on
$\mathcal{F}$ is also called the \emph{(Fock) vacuum} and denoted by $\Vacuum$.

\paragraph{CAR}
Associated with $\mathcal{F}$, there are natural linear, respectively anti-linear, maps
$c^*,c:\mathfrak{h}\to\mathcal{B}(\mathcal{F})$ called the \emph{creation-} and \emph{annihilation
operators} which are defined for $f\in\HilbertSpace$ and $\omega\in\FockSpace$ by
$c(\varphi)\doteq[c^*(\varphi)]^*$ and $c^*(f)\omega\doteq f\wedge\omega$, respectively. They
satisfy the \emph{canonical anti-commutation relations} (CAR)
\begin{equation}
  \begin{aligned}
    \left\{c^*(\varphi),c^*(\psi)\right\}=\left\{c(\varphi),c(\psi)\right\}&=0,&
    \left\{c^*(\varphi),c(\psi) \right\}=\SP{\varphi}{\psi},
    \quad\forall\varphi,\psi\in\mathfrak{h},
  \end{aligned}
  \label{fee10d27}
\end{equation}
and $c(\varphi)\Vacuum=0$ for all $\varphi\in\HilbertSpace$. The mappings
$c^*,c:\mathfrak{h}\to\mathcal{B}(\mathcal{F})$ induce a representation of the (abstract) CAR
algebra generated by $\mathfrak{h}$ \citep[see][Sec.~5.2.2]{Bratteli2002a}, called the \emph{Fock
representation}.

\paragraph{Density matrices}
We denote by $\mathcal{P}\doteq\mathcal{L}^1_+(\mathcal{F})\subseteq\mathcal{L}^2(\mathcal{F})$ the
cone of positive, trace-class operators on $\mathcal{F}$. Elements $\rho$ from the convex subset
$\mathcal{P}_1\subseteq\mathcal{P}$ which are \emph{normalized} in the sense that $\tr\{\rho\}=1$
are called \emph{density matrices on $\mathcal{F}$}. Elements of $\mathcal{P}_1$ uniquely represent
the \emph{normal states} on the C*-algebra $\mathcal{B}(\mathcal{F})$ \citep[see][Theorem
2.7]{Araki1999}.

\subsection{Finite-dimensional features}

We conclude this section by summarizing some more specific notions, which (partly) depend on the
finite-dimensionality of $\mathfrak{h}$.

\paragraph{Generalized creation- and annihilation operators}
By the CAR, we may extend $c,c^*$ to linear, respectively anti-linear, maps
$\Cd,\C:\mathcal{F}\to\mathcal{B}(\mathcal{F})$ via
\begin{equation}
    \label{90316b5f}
    \begin{aligned}
        \Cd(\omega)\eta&\doteq\omega\wedge\eta,&
        \C(\omega)&\doteq \left[\Cd(\omega)\right]^*.
    \end{aligned}
\end{equation}
Note that the definition of $\C$ is such that
$\C(\varphi_1\wedge\cdots\wedge\varphi_k)=c(\varphi_k)\cdots c(\varphi_1)$, for all
$\varphi_1,\dots\varphi_k\in\mathfrak{h}$. We call $\Cd,\C$ the \emph{generalized} creation- and
annihilation operators\footnote{This terminology is also used, e.g, in \cite{Stolarczyk2005}.}. Note
that the CAR \eqref{fee10d27} do \emph{not} hold for $\Cd$ and $\C$, when
$\varphi,\psi\in\mathfrak{h}$ are replaced by general $\omega,\eta\in\mathcal{F}$.

\paragraph{Polynomials in Creation- and Annihilation-Operators}
We are particularly interested in operators on $\mathcal{F}$, which are ``polynomials in creation-
and annihilation'' operators, i.e., elements in the complex $*$-subalgebra
$\mathcal{A}\subseteq\mathcal{B}(\mathcal{F})$ generated by
$\{c^*(\varphi)\mid\varphi\in\mathfrak{h}\}$. In the finite-dimensional case,
$\mathcal{A}=\mathcal{B}(\mathcal{F})$ \citep[see][Theorem 5.2.5]{Bratteli2002a} and we have a
natural linear map
\begin{equation}
  \Theta:\mathcal{F}\otimes\bar{\mathcal{F}}\ni
  \omega\otimes\bar{\eta}\mapsto\Cd(\omega)\C(\eta)\in\mathcal{A},
  \label{bfdb8cce}
\end{equation}
where $\bar{\mathcal{F}}$ denotes the \emph{conjugate} Hilbert space of $\mathcal{F}$
\citep[see][Sec.~1.2]{Derezinski2013}. In fact, by the Wick Theorem, $\Theta$ is surjective and
therefore an isomorphism, as the vector spaces involved are all finite-dimensional.

\paragraph{$k$-Body Operators}
Let $k\in\IN_0$. We call a sum of operators of the form $\Cd(\omega)\C(\eta)$ with
$\omega\in\mathcal{F}_r$, $\eta\in\mathcal{F}_s$ and $r+s=2k$ a \emph{$k$-particle operator}. More
generally, a sum of $l$-particle operators with $l\le k$ is called a \emph{$k$-body operator}, and we
denote the space of $k$-body operators by \kbOp. We also consider the $\IR$-subspace
$\kbOb\subseteq\kbOp$ of selfadjoint (or \emph{real}) elements of \kbOp, which are called
\emph{$k$-body observables}.
\begin{remark}[On the Terminology of $k$-Body Operators]\label{6c718f8b}
    There are different conventions regarding the notion of a \emph{$k$-body operator}. Especially
    in the physics literature this terminology usually refers to what we call a $k$-particle
    operator. For example, a typical Hamiltonian in second quantization is given by
    \eqref{46575df0}. In the physical literature, this operator would then often be considered as a
    sum of a one- and two-body operator, whereas in our convention \eqref{46575df0} is a sum of a one- and
    two-\emph{particle} operator and therefore a two-body operator.
\end{remark}

\paragraph{The Hilbert-Schmidt geometry}
Since in the finite-dimensional case we have $\mathcal{L}^2(\mathcal{F})=\mathcal{B}(\mathcal{F})$,
the mappings $\Theta$, $\Cd$ and $\C$ introduced above are in fact mappings between
(finite-dimensional) complex Hilbert spaces. In particular, using the natural isomorphism
$\mathcal{F}\otimes\bar{\mathcal{F}}\cong\mathcal{L}^2(\mathcal{F})$ the map $\Theta$ defined in
\eqref{bfdb8cce} gives rise to a linear automorphism
\begin{equation}
    \label{b4a2870e}
    \alpha:\mathcal{L}^2(\mathcal{F})\ni\left|\omega\rangle\langle\eta\right|
    \mapsto\Cd(\omega)\C(\eta)\in\mathcal{L}^2(\mathcal{F}).
\end{equation}

\section{Trace Formulas}
\label{9861ec75}

The goal of this section is to prove \thref{0aad}, which provides a formula for the Hilbert-Schmidt
inner product $\SP{a}{b}_\HS$ between certain monomials $a,b$ in creation and annhiliation
operators. Our approach is to evaluate
\begin{equation}\label{4440d370}
     \SP{a}{b}_\HS=\tr\{a^*b\}=\sum_I\SP{\varphi_I}{a^*b\varphi_I}_\FockSpace
\end{equation}
for a suitable basis $(\varphi_I)_I$ of $\FockSpace$ (\thref{6915af44}). The main work then is to
characterize the set $\mathfrak{M}$ of those $I$ with non-vanishing contributions in
\eqref{4440d370}
(\thref{d132a786}).

\subsection{Basic notation}
\paragraph{Set-theory}
For a set $X$, we denote by $\card{X}\in\IN\cup\{0,\infty\}$ the number of elements in $X$ and by
$\PowerSet{X}$ the system of all subsets of $X$. Given sets $A_1,\ldots,A_\Lambda\in\PowerSet{X}$,
we write $A_1\dotcup\cdots\dotcup A_\Lambda$ for their union $A_1\cup\cdots\cup A_\Lambda$ when we
want to indicate or require the $A_1,\ldots,A_\Lambda$ to be \emph{mutually disjoint}, i.e.,
$A_\alpha\cap A_\beta=\emptyset$ for all $1\le\alpha<\beta\le\Lambda$. Given a proposition $p$
(e.g., a set-theoretic relation like $x\in A\cap B)$ we write
\begin{equation}
    \mathbb{1}(p)\doteq\begin{cases}
        1&\text{if $p$ is true},\\
        0&\text{otherwise}.
    \end{cases}
\end{equation}
In the case where $p$ is of the form $a=b$, we also write $\delta_{a,b}$ for $\mathbb{1}(p)$ (the
\emph{Kronecker Delta}).

\paragraph{Orbital bases and induced Fock bases}
For the remainder of this paper, let $\mathfrak{h}$ be finite-dimensional, $\dim\mathfrak{h}\doteq
n<\infty$, and assume that $\{\varphi_1,\ldots,\varphi_n\}$ is a fixed orthonormal basis. Let
$\IN_n\doteq\{1,\ldots,n\}$ and $\PowerSet{\IN_n}$ be the family of subsets of $\IN_n$. For
$A=\{a_1,\cdots,a_k\}\subseteq\IN_n$ with $a_1<\cdots<a_k$ we define
\begin{equation}\label{fa55a342}
    \varphi_A\doteq
    \begin{cases}
        \varphi_{a_1}\wedge\cdots\wedge\varphi_{a_k}&A\ne\emptyset,\\
        \Vacuum&\text{for $A=\emptyset$.}
    \end{cases}
\end{equation}
Then, by definition \eqref{823d830d} of the inner product on $\FockSpace$,
$(\varphi_A)_{A\subseteq\IN_n}$ is an \emph{orthonormal} basis of $\FockSpace$ and, using Diracs
Bra-ket notation, $\left(\left|\varphi_A\rangle\langle\varphi_B\right|\right)_{A,B\subseteq\IN_n}$
is an orthonormal basis of $\HS$. Applying the generalized creation and annihilation operators, we
further define for $A,B\subseteq\IN_n$ the monomials
\begin{equation}
  \label{f76e1416}
  \begin{aligned}
    \Cd_A&\doteq \Cd(\varphi_A),&
    \C_A&\doteq \C(\varphi_A),&
    \C_{A,B}&\doteq \Cd_A\C_B,&
    \N_A&\doteq \C_{A,A}.
  \end{aligned}
\end{equation}

\subsection{Monomials acting on the induced Fock bases}
To efficiently deal with the signs occurring in computations with the monomials of the form
\eqref{f76e1416}, we introduce for $A_1,\ldots,A_k,B_1,\dots,B_l\subseteq\IN_n$ the \emph{multi-sign}
\begin{equation}
    \label{96423b67}
    \sign{A_1&\ldots&A_k}{B_1&\ldots&B_l}
    \doteq\SP{\varphi_{A_1}\wedge\cdots\wedge\varphi_{A_k}}{\varphi_{B_1}\wedge\cdots\wedge\varphi_{B_l}}.
\end{equation}
The main use of these multi-signs is to account for the signs occurring when reordering products of
elements of the form \eqref{fa55a342}, which is made precise by the following.
\begin{lemma}\label{e0b83582}
    The multi-sign \eqref{96423b67} vanishes, unless $A_1\dotcup\cdots\dotcup
    A_k=B_1\dotcup\cdots\dotcup B_l$. However, if $A_1\dotcup\cdots\dotcup
    A_k=B_1\dotcup\cdots\dotcup B_l$, then
    \begin{equation}
        \sign{A_1&\cdots&A_k}{B_1&\cdots&B_l}\left(\varphi_{A_1}\wedge\cdots\wedge\varphi_{A_k}\right)
        =\varphi_{B_1}\wedge\cdots\wedge\varphi_{B_l}.
    \end{equation}
    \begin{proof}
        Since the $\varphi_i$ anti-commute as elements in $\FockSpace$, its clear that
        $\varphi_{A_1}\wedge\cdots\wedge\varphi_{A_k}=0$ whenever the $A_i$ are not mutually
        disjoint (and similarly for the $B_i$). Therefore the right-hand side of \eqref{96423b67} trivially
        vanishes unless the $A_i$ and $B_i$ are mutually disjoint, respectively. Now consider the
        case where the $A_i$ and $B_i$ are mutually disjoint, but their unions $A$ respectively $B$
        are not equal, say there is $a\in A\setminus B$ for some $a\in\IN_n$. Then
        $\SP{\varphi_a}{\varphi_b}=0$ for all $b\in B$, thus
        $\SP{\varphi_A}{\varphi_B}=0$ by definition \eqref{823d830d} and
        \begin{equation}
            \sign{A_1&\cdots&A_k}{B_1&\cdots&B_l}=\pm\SP{\varphi_A}{\varphi_B}=0,
        \end{equation}
        which proves the first part. For the second part, assume that $A_1\dotcup\cdots\dotcup A_k=B_1\dotcup\cdots\dotcup B_l$. Then, by anti-commuting
        the $\varphi_i$, there is $\lambda\in\{-1,+1\}$ such that
        \begin{equation}
            \varphi\doteq\varphi_{A_1}\wedge\cdots\wedge\varphi_{A_k}
            =\lambda\cdot\varphi_{B_1}\wedge\cdots\wedge\varphi_{B_l}\doteq\lambda\cdot\tilde\varphi
        \end{equation}
        Using the same argument, we find that $\tilde\varphi=\pm\varphi_A$, thus
        $\|\tilde\varphi\|^2=1$. Consequently,
        \begin{equation}
            \begin{split}
                \sign{A_1&\cdots&A_k}{B_1&\cdots&B_l}\varphi_{A_1}\wedge\cdots\wedge\varphi_{A_k}&=
                \SP{\varphi}{\tilde\varphi}\varphi=\lambda^2\|\tilde\varphi\|^2\tilde\varphi=\tilde\varphi
                \\&=\varphi_{B_1}\wedge\cdots\wedge\varphi_{B_l}.
            \end{split}
        \end{equation}
    \end{proof}
\end{lemma}

\begin{lemma}\label{35f7b9dd}
  For $A,B,I\subseteq\IN_n$ we have
  \begin{align}
      \Cd_A\varphi_I\;&=\;\one(A\cap I=\emptyset)\sign{A&I}{A\cup I}\varphi_{A\cup I}\label{2fa63a0c}\\
      \C_A\varphi_I\;&=\;\mathbb{1}(A\subseteq I)\sign{A&I\setminus A}{I}\varphi_{I\setminus A}\label{17922d73}.
  \end{align}
  \begin{proof}
    If $A\cap I\ne\emptyset$ then $\Cd_A\varphi_I=0$ and also the right hand side of \eqref{2fa63a0c}
    vanishes due to \thref{e0b83582}. Otherwise, if $A\cap I=\emptyset$ then \thref{e0b83582} implies
    \begin{equation}
        \Cd_A\varphi_I=\varphi_A\wedge\varphi_I=\sign{A&B}{A\cup B}\varphi_{A\cup B},
    \end{equation}
    which completes the proof of \eqref{2fa63a0c}.

    To prove \eqref{17922d73} note that, since $(\varphi_J)_{J\subseteq\IN_n}$ is an orthonormal basis
    of $\FockSpace$, we have
    \begin{equation}
        \label{f5b50e17}
        \C_A\varphi_I=\sum_{J\subseteq\IN_n}\SP{\C_A\varphi_I}{\varphi_J}\varphi_J.
    \end{equation}
    Unwinding the definitions and using \thref{e0b83582}, we compute
    \begin{equation}
        \label{9d211342}
        \begin{split}
            \SP{\C_A\varphi_I}{\varphi_J}\varphi_J
            &=\SP{\varphi_I}{\varphi_A\wedge\varphi_J}=\sign{I}{A&J}\\
            &=\one(A\subseteq I)\one(J=A\setminus I)\sign{I}{A&I\setminus A}.
        \end{split}
    \end{equation}
    thus \eqref{17922d73} follows by combining \eqref{f5b50e17} and \eqref{9d211342}.
  \end{proof}
\end{lemma}
\begin{remark}
    Definition \eqref{fa55a342} of the Fock space basis elements $\varphi_A$ naturally generalizes
    to the case where $A$ is a \emph{string} over the alphabet $\IN_n$. Within this generalized
    framework, the multi-sign \eqref{96423b67} can be interpreted as the anti-symmetric Kronecker
    Delta (see, e.g., the ``algebraic framework'' in \cite{Stolarczyk2010}).
\end{remark}

\subsection{Derivation of the trace formula}\label{614830ea}
\begin{proposition}\label{6915af44}
  Let $A,B,C,D\subseteq\IN_n$, then
  \begin{equation}
    \label{ba99c7e1}
    \SP{\C_{A,B}}{\C_{C,D}}_{\HS}=
    \sum_{I\in\mathfrak{M}}
    \sign{A&I\setminus B}{C&I\setminus D}\sign{I}{B&I\setminus B}\sign{I}{D&I\setminus D}
  \end{equation}
  where $\mathfrak{M}\doteq\mathfrak{M}(A,B,C,D)$ is the family of all $I\subseteq\IN_n$ such that
  \begin{enumerate}
      \item $B\cup D\subseteq I$ and
      \item $A\dotcup(I\setminus B)=C\dotcup(I\setminus D)$.
  \end{enumerate}
  \begin{proof}
    Since $(\varphi_I)_{I\subseteq\IN_n}$ is an orthonormal basis of $\FockSpace$, we have
    \begin{equation}
        \label{dd29cf3a}
        \SP{\C_{A,B}}{\C_{C,D}}=\tr\{\Cd_B\C_A\Cd_C\C_D\}
        =\sum_{I\subseteq\IN_n}\SP{\Cd_A\C_B\varphi_I}{\Cd_C\C_D\varphi_I}
    \end{equation}
    Using \thref{35f7b9dd}, we compute for arbitrary $I\subseteq\IN_n$
    \begin{equation}
      \label{52c1da22}
      \begin{aligned}
        \C_{A,B}\varphi_I&=\Cd_A(\C_B\varphi_I)
        =\mathbb{1}(B\subseteq I)\sign{I}{B&I\setminus B}
        \Cd_A \varphi_{I\setminus B}\\
        &=\mathbb{1}(B\subseteq I)\mathbb{1}(A\cap(I\setminus B)=\emptyset)
        \sign{I}{B&I\setminus B}\varphi_{A}\wedge\varphi_{I\setminus B},
      \end{aligned}
    \end{equation}
    and similarly for $\C_{C,D}\varphi_I$, which yields
    \begin{equation}
        \label{eb863a23}
        \SP{\C_{A,B}\varphi_I}{\C_{C,D}\varphi_I}
        =\mathbb{1}(I\in\mathfrak{M})
        \sign{A&I\setminus B}{C&I\setminus D}
        \sign{I}{B&I\setminus B}
        \sign{I}{D&I\setminus D}.
    \end{equation}
    Combining \eqref{eb863a23} with \eqref{dd29cf3a}, the assertion follows.
  \end{proof}
\end{proposition}

As stated in \thref{6915af44}, the contributing sets $I\subseteq\IN_n$ in \eqref{ba99c7e1} must
satisfy certain set-theoretic compatibility relations with the given sets $A,B,C$ and $D$. Moreover,
\thref{6915af44} is of limited use because of the complicated signs occuring in \eqref{ba99c7e1}.
The main part of this paper therefore is to overcome these difficulties by a careful analysis of the
set $\mathfrak{M}$ of contributing subsets $I\subseteq\IN_n$.

\begin{proposition}\label{d132a786}
  Let $\mathfrak{M}=\mathfrak{M}(A,B,C,D)$ as in \thref{6915af44}. Then the following conditions are
 equivalent:
  \begin{enumerate}
    \item $\mathfrak{M}\ne\emptyset$, \label{d132a786-i1}
    \item $A\dotcup(D\setminus B)=C\dotcup(B\setminus D)$, \label{d132a786-i2}
    \item $B\cup D\in\mathfrak{M}$, \label{d132a786-i3}
    \item $A\setminus B=C\setminus D$ and $B\setminus A=D\setminus C$.
      \label{d132a786-i4}
  \end{enumerate}
  In any of these cases,
  \begin{equation}
    \mathfrak{M}=\{(B\cup D)\dotcup N\mid N\cap(A\cup C)=\emptyset\}.
    \label{0af28023}
  \end{equation}
  \begin{proof}
    We will first show the equivalence of the conditions \ref{d132a786-i1}-\ref{d132a786-i3}. The
    equivalence of \ref{d132a786-i2} and \ref{d132a786-i4} follows from a purely set-theoretic
    argument, see \thref{2cc8} below.

    \ref{d132a786-i1}$\Rightarrow$\ref{d132a786-i2}: Choose $M\in\mathfrak{M}$.  By definition of
    $\mathfrak{M}$, $B\cup D\subseteq M$, we may write $M=(B\cup D)\dotcup N$ so that $M\setminus
    B=(D\setminus B)\dotcup N$.  Since $A\cap(M\setminus B)=\emptyset$ by definition of
    $\mathfrak{M}$, also $A\cap(D\setminus B)\subseteq A\cap(M\setminus B)=\emptyset$, and similarly
    $C\cap(B\setminus D)=\emptyset$. Moreover, we have $A\cap N\subseteq A\cap((D\setminus B))\cup
    N)=A\cap(M\setminus B)=\emptyset$ and similarly $C\cap N=\emptyset$. In summary, we have
    $\left(A\cup(D\setminus B) \right)\dotcup N=A\cup(M\setminus B)=C\cup(M\setminus D)
    =\left(C\cup(B\setminus D) \right)\dotcup N$ and therefore $A\cup (D\setminus B)=C\cup
    (B\setminus D)$.

    \ref{d132a786-i2}$\Rightarrow$\ref{d132a786-i3}: By definition of $\mathfrak{M}$, $M\doteq B\cup
    D\in\mathfrak{M}$ if and only if $A\dotcup(M\setminus B)=C\dotcup(M\setminus D)$, but by
    construction $M\setminus B=D\setminus B$ and $M\setminus D=B\setminus D$.

    \ref{d132a786-i3}$\Rightarrow$\ref{d132a786-i1}: this follows trivially.

    Now it remains to prove \eqref{0af28023}, given the
    conditions \ref{d132a786-i1}-\ref{d132a786-i4} hold. Denote the right-hand side of
    \eqref{0af28023} by $\tilde{\mathfrak{M}}$.

    $\mathfrak{M}\subseteq\tilde{\mathfrak{M}}$: Choose some $M\in\mathfrak{M}$. Since $B\cup
    D\subseteq M$, we can write $M=(B\cup D)\dotcup N$ for some $N\subseteq I\setminus (B\cup D)$
    and now need to show that $N\cap(A\cup C)=\emptyset$.  Since $A\cap(M\setminus B)=\emptyset$ by
    definition of $\mathfrak{M}$, also $A\cap(D\setminus B)\subseteq A\cap(M\setminus B)=\emptyset$,
    and similarly $C\cap(B\setminus D)=\emptyset$. Moreover, we have $A\cap N\subseteq
    A\cap((D\setminus B))\cup N)=A\cap(M\setminus B)=\emptyset$ and similarly $C\cap N=\emptyset$,
    thus $N\cap(A\cup C)=\emptyset$.

    $\tilde{\mathfrak{M}}\subseteq\mathfrak{M}$: Let $M\doteq(B\cup D)\dotcup
    N\in\tilde{\mathfrak{M}}$, i.e., $N\cap(A\cup C)=\emptyset$. Clearly, $B\cup D\subseteq M$.
    Moreover, by assumption we have $A\dotcup(D\setminus B)=C\dotcup(B\setminus D)$, thus
    \begin{equation}
      A\cap(M\setminus B)=A\cap((D\setminus B\cup N)
      =(A\cap(D\setminus B))\cup (A\cap N)=\emptyset.
    \end{equation}
    Similarly, $C\cap(M\setminus D)=\emptyset$. Finally,
    \begin{equation}
      \begin{aligned}
        A\cup(M\setminus B)&=A\cup((D\setminus B)\cup N)=(A\cup(D\setminus B))\cup N\\
        &=(C\cup(B\setminus D))\cup N=C\cup(M\setminus D),
      \end{aligned}
    \end{equation}
    thus $M\in\mathfrak{M}$, which completes the proof.
  \end{proof}
\end{proposition}

\begin{lemma}\label{2cc8}
  Let $X$ be a set and $A,B,C,D\subseteq X$. Then the following conditions are equivalent
  \begin{enumerate}
    \item $A\dotcup(D\setminus B)=C\dotcup(B\setminus D)$, \label{d078}
    \item $A\setminus B=C\setminus D$ and $B\setminus A=D\setminus C$. \label{ef56}
  \end{enumerate}
  \begin{proof}
    \ref{d078}$\Rightarrow$\ref{ef56}: Let $x\in A\setminus B$. Then $x\in A\subseteq
    A\dotcup(D\setminus B) =C\dotcup(B\setminus D)$, thus $x\in C$. Moreover, since $(A\setminus
    B)\cap D=A\cap(D\setminus B)=\emptyset$, we have $x\not\in D$, hence $x\in C\setminus D$. This
    shows that $A\setminus B\subseteq C\setminus D$. Exchanging the roles of $A,C$ and $B,D$
    respectively, also $C\setminus D\subseteq A\setminus B$.

    Moreover, let $x\in B\setminus A$. If $x\not\in D$ then $x\in B\setminus D\subseteq
    C\dotcup(B\setminus D)=A\dotcup(D\setminus B)$, i.e.,  $x\in A$, contradicting our assumption
    $x\in B\setminus A$.  Hence, $x\in D$. Also, if $x\in C$ then $x\in C\dotcup(B\setminus D)
    =A\dotcup(D\setminus B)$, so $x\in D\setminus B$, which contradicts $x\in B$, hence $x\not\in
    C$. This shows $B\setminus A\subseteq D\setminus C$.  Again, by renaming $A,B,C$ and $D$, we
    also see $D\setminus C\subseteq B\setminus A$.

    \ref{ef56}$\Rightarrow$\ref{d078}: We compute
    \begin{equation}
      \begin{split}
        A\cap(D\setminus B)
        &=A\cap D\cap B^c=(A\setminus B)\cap D=(C\setminus D)\cap D=\emptyset.
      \end{split}
      \label{c52e}
    \end{equation}
    Exchanging the roles of $A,C$ and $B,D$, we also get $C\cap(B\setminus D)=\emptyset$. 
    To show that $A\cup (D\setminus B)=C\cup(B\setminus D)$, first note that
    \begin{equation}
      \begin{aligned}
        A\cap D^c&
        =(A\cap D^c\cap B)\cup(A\cap D^c\cap B^c)
        \subseteq (B\setminus D)\cup(A\setminus B)\\
        &=(B\setminus D)\cup(C\setminus D)
        \subseteq C\cup(B\setminus D)
      \end{aligned}
      \label{38a8}
    \end{equation}
    and
    \begin{equation}
      \begin{aligned}
        A\cap B&
        =A\cap(A\cap B)
        \subseteq A\cap(B\setminus A)^c
        =A\cap(D\setminus C)^c
        =A\cap(C\cup D^c)\\
        &=(A\cap C)\cup(A\cap D^c)
        \subseteq C\cup B\setminus D,
      \end{aligned}
      \label{c8b5}
    \end{equation}
    where we used \eqref{38a8} in the last step. Consequently, we conclude
    \begin{equation}
      A
      \stackrel{\eqref{c52e}}{\subseteq} A\cap(D\setminus B)^c
      =A\cap(D^c\cup B)
      =(A\cap D^c)\cup(A\cap B)
      \subseteq C\cup(B\setminus D),
      \label{3eb8}
    \end{equation}
    where we used \eqref{38a8} and \eqref{c8b5} in the last step.
    Moreover, we have
    \begin{equation}
      \begin{aligned}
        D\setminus B
        &\stackrel{\eqref{c52e}}{\subseteq} (D\setminus B)\cap A^c
        =[(D\setminus B)\cap A^c\cap C[\cup[(D\setminus B)\cap A^c\cap C^c]\\
        &\subseteq C\cup(D\cap C^c\cap A^c)=C\cup(B\cap A^c)
        \subseteq C\cup B,
      \end{aligned}
      \label{4382}
    \end{equation}
    and intersecting both sides of this inclusion with $B^c$, we obtain $D\setminus B\subseteq
    C\setminus B\subseteq C$. Combined with \eqref{3eb8}, this shows $A\cup (D\setminus B)\subseteq
    C\cup(B\setminus D)$ and, by exchanging the roles of $A,C$ and $B,D$, the converse inclusion
    follows as well.
  \end{proof}
\end{lemma}

\begin{remark}
  Lemma \ref{2cc8} can be further generalized by noting that the given conditions are also equivalent
  to the following (equivalent) conditions:
  \begin{enumerate}
    \item $B\setminus D=A\setminus C$ and $D\setminus B=C\setminus A$,
    \item $B\dotcup(A\setminus C)=D\dotcup(C\setminus A)$.
  \end{enumerate}
\end{remark}

\begin{proposition}[Trace Formula]\label{0aad}
  Let $K,A,B\subseteq\IN_n$ and $L,C,D\subseteq\IN_n$ be mutually disjoint, respectively. Then
  \begin{equation}\label{fd5f33fe}
    \SP{\N_K\C_{A,B}}{\N_L\C_{C,D}}_{\HS}=\delta_{A,C}\delta_{B,D}
    \cdot 2^{n-\card{A\cup B\cup K\cup L}}.
  \end{equation}
  \begin{proof}
    Using \thref{e0b83582} and \thref{35f7b9dd}, we find for any $I\subseteq\IN_n$
    \begin{equation}
    \begin{aligned}
      \N_K\varphi_I&=\Cd_K\left(\C_K\varphi_I\right)
      =\mathbb{1}(K\subseteq I)\sign{I}{K&I\setminus K}
      \Cd_K\varphi_{I\setminus K}\\
      &=\mathbb{1}(K\subseteq I)\sign{I}{K&I\setminus K}\varphi_{K}\wedge\varphi_{I\setminus K}
      =\mathbb{1}(K\subseteq I)\varphi_I.
    \end{aligned}
    \end{equation}
    Combined with \thref{35f7b9dd}, we therefore get for any $I\subseteq\IN_n$
    \begin{equation}
      \begin{aligned}
        \N_K\C_{A,B}\varphi_I&=
        \mathbb{1}(K\subseteq A\cup (I\setminus B))\mathbb{1}(B\subseteq I)\mathbb{1}(A\cap I\setminus B=\emptyset)\\
        &\qquad\cdot\sign{I}{B&I\setminus B}\varphi_A\wedge\varphi_{I\setminus B}.
      \end{aligned}
      \label{340b}
    \end{equation}
    Consequently, we have with $\mathfrak{M}=\mathfrak{M}(A,B,C,D)$
    as in \thref{d132a786}
    \begin{multline}
      \SP{\N_K\C_{A,B}\varphi_I}{\N_L\C_{C,D}\varphi_I}=
      \mathbb{1}(I\in\mathfrak{M})
      \mathbb{1}[K\subseteq A\cup(I\setminus B)]
      \mathbb{1}[L\subseteq C\cup(I\setminus D)]\\
      \cdot\sign{A&I\setminus B}{C&I\setminus D}\sign{I}{B&I\setminus B}\sign{I}{D&I\setminus D}
      \label{750c}
    \end{multline}
    Since $A\cap B=C\cap D=\emptyset$ by assumption, \thref{d132a786} implies that
    $\mathbb{1}(I\in\mathfrak{M})=\delta_{A,C}\delta_{B,D}\mathbb{1}(B\subseteq I)
    \mathbb{1}(I\cap A=\emptyset)$. Thus \eqref{750c} equals
    \begin{equation}
      \delta_{A,C}\delta_{B,D}
      \mathbb{1}(B\subseteq I)\mathbb{1}(I\cap A=\emptyset)
      \mathbb{1}[K\cup L\subseteq A\cup (I\setminus B)].
      \label{457c8d8f}
    \end{equation}
    Now observe that for $A=C$ we have $L\cap A=L\cap C=\emptyset$, i.e.,
    $K\cup L\subseteq A\cup(I\setminus B)$ is equivalent to $K\cup L\subseteq I\setminus B$, which
    is further equivalent to $K\cup L\subseteq I$. Hence
    \eqref{750c} equals
    \begin{equation}
      \delta_{A,C}\delta_{B,D}
      \mathbb{1}(I\cap A=\emptyset)
      \mathbb{1}(B\cup K\cup L\subseteq I)
      \label{142c120a}
    \end{equation}
    and, by summing \eqref{142c120a} over all $I\subseteq\IN_n$, we find
    \begin{equation}
      \SP{\N_K\C_{A,B}}{\N_L\C_{C,D}}=
      \delta_{A,C}\delta_{B,D}\card{\mathfrak{P}[\IN_n\setminus(A\cup
    B\cup K\cup L)]}.
    \end{equation}
  \end{proof}
\end{proposition}
\begin{example}[Trace of the Particle Number Operator]\label{3dea59d9}
    Let $\dim\mathfrak{h}=n<\infty$. By \thref{35f7b9dd}, the \emph{particle number operator}
    $\hat{\IN}\doteq\sum_{i=1}^n n_i$ can be written as $\hat{\IN}=\bigoplus_{k=0}^n
    k\cdot\operatorname{id}_{\Lambda^k\mathfrak{h}}$. Consequently, its trace is given by
    $\sum_{k=0}^n k\cdot\binom{n}{k}$. On the other hand, \thref{0aad} implies
    $\tr\{\hat{\IN}\}=\sum_{i=1}^n\SP{\mathbb{1}}{n_i}=n\cdot 2^{n-1}$. Thus we proved
    the well-known identity
    \begin{equation}
        \sum_{k=0}^nk\binom{n}{k}=\tr\{\hat{\IN}\}=n\cdot 2^{n-1},
    \end{equation}
    which also follows from differentiating $(1+x)^n$ with respect to $x$ and evaluating at $x=1$.
\end{example}

\section{Orthonormalization}\label{280ac894}
In this section, given an orthonormal basis in $\HilbertSpace$, we will construct explicit
orthogonal bases of $\HS$ which restrict to the spaces of $k$-body operators and $k$-body
observables, respectively.

\subsection{Orthonormal basis of \texorpdfstring{$\HS$}{L2(F)}}
\label{3f89dfe9}
As implied by \thref{0aad}, the monomials $(\N_K)_{K\subseteq\IN_n}$ are \emph{not} pairwise
orthogonal. Inspired by computer algebraic experiments using Gram-Schmidt orthogonalization in
low-dimensional cases, we introduce for $K\subseteq\IN_n$ the element
\begin{equation}
  b_K\doteq\sum_{I\subseteq K}(-2)^\card{I}\N_I\in\HS.
  \label{55e8df1c}
\end{equation}
As we will see in \thref{9faf7dc5}, the $b_K$ are pairwise orthogonal and can be used to
construct an orthogonal basis of $\HS$. The key ingredient is the following lemma, which is
essentially a consequence of the binomial formula.
\begin{lemma}\label{6b855753}
  Let $K,L$ be finite sets. Then
  \begin{equation}
    \sum_{I\subseteq K}\sum_{J\subseteq L}(-2)^{\card{I}+\card{J}}2^{-\card{I\cup
    J}}=\delta_{KL}.
  \end{equation}
  \begin{proof}
    Let $M\doteq K\cap L$. We compute
    \begin{equation}
      \label{12998d2d}
      \begin{aligned}
        S\doteq\sum_{\substack{I\subseteq K\\J\subseteq L}}(-2)^{\card{I}+\card{J}}2^{-\card{I\cup J}}
        =\sum_{\substack{I\subseteq K\\J\subseteq L}}\frac{(-1)^{\card{I}+\card{J}}}{2^{-\card{I\cap J}}},
      \end{aligned}
    \end{equation}
    where we have used that $\card{I\cup J}=\card{I}+\card{J}-\card{I\cap J}$.
    Since every $I\subseteq K$ can be written uniquely as $I=I_1\dotcup I_2$ with
    $I_1\doteq(I\cap M)\subseteq M$ and $I_2\doteq I\setminus I_1\subseteq K\setminus
    M$ and (similarly for $J\subseteq L$), we find
    \begin{equation}
      \label{176b702e}
      \begin{aligned}
        S=\sum_{I_1,J_1\subseteq M}\frac{(-1)^{\card{I_1}+\card{J_1}}}{2^{-\card{I_1\cap J_1}}}
        \sum_{I_2\subseteq K\setminus M}(-1)^{\card{I_2}}
        \sum_{J_2\subseteq K\setminus M}(-1)^{\card{J_2}}.
      \end{aligned}
    \end{equation}

    By the binomial formula, for any finite set $X$ and $a\in\IC$ we have
    \begin{equation}
      \sum_{Y\subseteq X}a^{\card{Y}}=(1+a)^{\card{X}}.
      \label{4523374f}
    \end{equation}
    In particular, for $a=-1$ we have
    $\sum_{Y\subseteq X}(-1)^{\card{Y}}=\mathbb{1}(X=\emptyset)$. Hence
    \begin{equation}
      \label{4767b0ce}
      \begin{aligned}
        \sum_{I_2\subseteq K\setminus M}(-1)^{\card{I_2}}
        \sum_{J_2\subseteq L\setminus M}(-1)^{\card{J_2}}
        &=\mathbb{1}(K\setminus M=\emptyset)\mathbb{1}(L\setminus M=\emptyset)\\
        &=\mathbb{1}(K\subseteq L)\mathbb{1}(L\subseteq K)=\delta_{KL}.
      \end{aligned}
    \end{equation}
    Inserting \eqref{4767b0ce} in \eqref{176b702e}, we find
    \begin{equation}
      \label{fa9311c0}
      \begin{aligned}
        S&=
        \delta_{KL}\sum_{I,J\subseteq M}
        \frac{(-1)^{\card{I}+\card{J}}}{2^{-\card{I\cap J}}}.
      \end{aligned}
    \end{equation}
    To evaluate the sum in \eqref{fa9311c0}, instead of summing over all $I,J\subseteq M$,
    we sum over all $X\doteq I\cap J\subseteq M$, $I_3\doteq I\setminus X\subseteq M\setminus X$ and
    $J_3\doteq J\setminus(X\dotcup I_3)\subseteq M\setminus(X\dotcup I_3)$ and apply
    \eqref{4523374f} once
    again:
    \begin{equation}
      \label{b22aac60}
      \begin{aligned}
        \sum_{\substack{I\subseteq M\\J\subseteq M}}
        \frac{(-1)^{\card{I}+\card{J}}}{2^{-\card{I\cap J}}}
        &=\sum_{X\subseteq M}2^{\card{X}}\sum_{I_3\subseteq M\setminus X}(-1)^{\card{I_3}}
        \sum_{J_3\subseteq M\setminus(X\dotcup I_3)}(-1)^{\card{J_3}}\\
        &=\sum_{X\subseteq M}2^{\card{X}}\sum_{I_3\subseteq M\setminus X}(-1)^{\card{I_3}}
        \mathbb{1}(I_3=M\setminus X)\\
        &=\sum_{X\subseteq M}2^{\card{X}}(-1)^{\card{M\setminus X}}
        =(-1)^{\card{M}}\sum_{X\subseteq M}(-2)^{\card{X}}\\
        &=(-1)^{\card{M}}(-1)^{\card{M}}=1.
      \end{aligned}
    \end{equation}
    Combining \eqref{fa9311c0} and \eqref{b22aac60}, the assertion follows.
  \end{proof}
\end{lemma}

\begin{theorem}\label{9faf7dc5}
  Let $b_K$ be defined as in \eqref{55e8df1c}, then an orthonormal basis of $\HS$ is explicitly
  given by
  \begin{equation}\label{2ee45bf4}
    \HSbasis=\left\{\left.\frac{b_K\C_{I,J}}{\sqrt{2^{n-\card{I\dotcup J}}}}\in\HS\right|
        K,I,J\subset\IN_n\text{ pairwise disjoint}\right\}.
  \end{equation}
  \begin{proof}
    Let $K,A,B\subseteq\IN_n$ and $L,C,D\subseteq\IN_n$ be mutually disjoint, respectively. By
    definition of $b_K$ and using \thref{0aad}, we obtain
    \begin{equation}\label{7ed31441}
      \begin{aligned}
        \SP{b_K\C_{A,B}}{b_L\C_{C,D}}&=
        \sum_{I\subseteq K}\sum_{J\subseteq L}(-2)^{\card{I}+\card{J}}
        \SP{\N_I\C_{A,B}}{\N_J\C_{C,D}}\\
        &=\sum_{I\subseteq K}\sum_{J\subseteq L}(-2)^{\card{I}+\card{J}}
        \delta_{AC}\delta_{BD}2^{n-\card{(A\dotcup B)\cup(I\cup J)}}\\
        &=\delta_{AC}\delta_{BD}2^{n-\card{A\dotcup B}}
        \left(\sum_{I\subseteq K}\sum_{J\subseteq L}(-2)^{\card{I}+\card{J}}
        2^{-\card{I\cup J}}\right)\\
        &=\delta_{AC}\delta_{BD}2^{n-\card{A\dotcup B}}\delta_{KL},
      \end{aligned}
    \end{equation}
    where we used that for $A=C, B=D, I\subseteq K$ and $J\subseteq L$ we have $\card{A\cup
    B\cup I\cup J}=\card{A\cup B}+\card{I\cup J}$ in the third step and \thref{6b855753} (see below)
    in the last step. This shows that \eqref{2ee45bf4} is an orthonormal basis of its span $S$.
    Noting that
    \begin{equation}
      \dim S=\card{\HSbasis}=
      \card{\{f:\IN_n\to\{1,2,3,4\}\}}
      =4^n=\dim\HS,
    \end{equation}
    we conclude that $S=\HS$.
  \end{proof}
\end{theorem}

\subsection{Orthonormal basis of $k$-body operators}
Having established $\HSbasis$ as an orthonormal basis of $\HS$, we now proceed and show that
$\HSbasis$ restricts to a basis of $\kbOp$ for all $k\in\IN_0$ (\thref{46f8bbb2}).

\begin{lemma}\label{d29294a2}
    A basis of $\kbOp$ is explicitly given by
    \begin{equation}
        \label{008214dd}
        \mathfrak{B}_0\doteq\left\{\C_{I,J}\left|I,J\subseteq\IN_n,\card{I}+\card{J}=2l
        \text{ with }0\le l\le k\right.\right\},
    \end{equation}
    in particular, we have $\dim_\IC\kbOp=\sum_{l=0}^k\binom{2n}{2l}$.
    \begin{proof}
        Since the mapping $\alpha$ defined in \eqref{b4a2870e} is a linear automorphism of $\HS$,
        the $\C_{I,J}=\alpha\left(\left|\varphi_I\rangle\langle\varphi_J\right|\right)$ with
        $I,J\subseteq\IN_n$ form a basis of $\HS$. An element $A\in\HS$ of the form
        \begin{equation}
          A=\sum_{I,J\subseteq\IN_n}A_{I,J}\C_{I,J}
        \end{equation}
        is a $k$-body operator if and only if $A_{I,J}=0$ whenever $\card{I}+\card{J}$ is odd or
        $\card{I}+\card{J}>2k$. In other words, \eqref{008214dd} a basis of $\kbOp$ and
        \begin{equation}
            \dim_\IC\kbOp=\card{\mathfrak{B}_0}=\sum_{l=0}^k\sum_{i=0}^{2l}\binom{n}{i}\binom{n}{2l-i}=
            \sum_{l=0}^k\binom{2n}{2l},
        \end{equation}
        where we used Vandermonde's identity.
    \end{proof}
\end{lemma}

\begin{theorem}\label{46f8bbb2}
    The orthonormal $\IC$-basis $\HSbasis$ of $\HS$ given in \thref{9faf7dc5} restricts to an
    orthonormal basis $\kbOpBasis$ of the space $\kbOp$ of $k$-body operators. More specifically, we
    have
    \begin{equation}
        \label{836a5fd6}
        \kbOpBasis\doteq\HSbasis\cap\kbOp
        =\left\{\frac{b_K\C_{I,J}}{\sqrt{2^{n-\card{I\cup J}}}}
        \left|\begin{gathered}
            K,I,J\subset\IN_n\text{ pairwise disjoint,}\\
            \card{I}+\card{J}+2\card{K}=2l\text{ with }0\le l\le k
        \end{gathered}\right.\right\}.
    \end{equation}
    \begin{proof}
        Let $b\in\HSbasis$, i.e.,
        \begin{equation}
            b=b_K\C_{I,J}=\sum_{L\subseteq K}\frac{(-2)^\card{L}}{\sqrt{2^{n-\card{I\cup J}}}}n_L\C_{I,J}
        \end{equation}
        for $K,I,J\subseteq\IN_n$ pairwise disjoint. Since $n_L\C_{I,J}=\pm\C_{I\dotcup L,J\dotcup
        L}$ for every $L\subseteq K$, \thref{d29294a2} implies that $b\in\kbOp$ if and only if
        $\card{I}+\card{J}+2\card{K}=2l$ for some $0\le l\le k$, which proves \eqref{836a5fd6}.
        Finally, noting that we have a bijection $\HSbasis\ni b_K\C_{I,J}\to\C_{I\dotcup
        K,J\dotcup K}\in\mathfrak{B}_0$ with inverse $\C_{I,J}\mapsto b_{I\cap J}\C_{I\setminus
        J,J\setminus I}$, we conclude that $\card{\kbOpBasis}=\card{\mathfrak{B}_0}=\dim\kbOp$ and
        therefore $\HSbasis_k$ is a basis of $\kbOp$.
    \end{proof}
\end{theorem}

\subsection{Orthonormal basis of $k$-body observables}
The orthonormal $\IC$-basis $\HSbasis$ of $\HS$ as given in \thref{9faf7dc5} does not
immediately restrict to bases of $k$-body \emph{observables}, since $\mathfrak{B}_\IC$ contains
elements which are not self-adjoint. For example, if $I\subset\IN_n$ is non-empty, then
\[\left(b_\emptyset\C_{I,\emptyset}\right)^*=\C_I\ne\Cd_I=b_\emptyset\C_{I,\emptyset}.\]However,
$\mathfrak{B}_\IC$ has the special property that $\mathfrak{B}_\IC=\{b^*\mid
b\in\mathfrak{B}_\IC\}$, which allows us to obtain an orthonormal basis of self-adjoint elements by
a suitable unitary transformation of $\HS$. The general principle of this idea
is given by the following.

\begin{lemma}\label{21a201d6}
    Let $\mathcal{H}$ be a finite-dimensional, complex Hilbert space with real structure $J$ and
    $\mathfrak{B}$ an orthonormal $\IC$-basis with $J(\mathfrak{B})\subseteq\mathfrak{B}$. Then
    \begin{enumerate}
        \item \label{21a201d6-i1} $\mathfrak{B}$ is of the form
        \begin{equation}\label{291efc26}
            \mathfrak{B}=(a_1,\ldots,a_k,b_1,b_1^*,\ldots,b_l,b_l^*)\text{ with }a_i=a_i^*
            \quad\forall 1\le
            i\le k.
        \end{equation}
        \item \label{21a201d6-i2} An orthonormal $\IR$-basis of $V_\IR\doteq\{v\in V\mid J(v)=v\}$
        is given by
        \begin{equation}
            \mathfrak{B}_\IR\doteq\left(a_1,\ldots,a_k,
            \sqrt{2}\Re(b_1),\sqrt{2}\Im(b_1),
            \ldots,\sqrt{2}\Re(b_l),\sqrt{2}\Im(b_l)\right)
        \end{equation}
    \end{enumerate}
    [Here, $\Re(a)\doteq\frac{1}{2}(a+a^*)$ and $\Im(a)\doteq\frac{1}{2i}(a-a^*))$ denote the real-
    and imaginary part of $a$, respectively]
    \begin{proof}
    \ref{21a201d6-i1} Since $J(\mathfrak{B})\subseteq\mathfrak{B}$ and $J^2=1$, $J$ defines an
    action of $\IZ/2\IZ$ on $\mathfrak{B}$. The set $\mathfrak{B}$ is decomposed into the orbits of
    this action, which are either of length $1$ or length $2$ by the orbit-stabilizer Theorem. By
    construction, the orbits of length $1$ are of the form $\{a=a^*\}$ and the orbits of length $2$
    are of the form $\{b,b^*\}$, hence the desired form \eqref{291efc26} is obtained by selecting an
    element in each orbit of $\mathfrak{B}$.

    \ref{21a201d6-i2} Let $f:V\to V$ be the $\IC$-linear map mapping $\mathfrak{B}$ to
    $\mathfrak{B}_\IR$. Then $f$ is represented with respect to $\mathfrak{B}$ by the unitary matrix
    \begin{equation}
        \one_k\oplus \underbrace{U\oplus\cdots\oplus U}_{l\text{ times}}
        \quad\text{with}\quad U\doteq
        \frac{1}{\sqrt{2}}\begin{pmatrix}
            1 & -i\\
            1 & i
        \end{pmatrix}\in U(2).
    \end{equation}
    In particular, with $\mathfrak{B}$ also $\mathfrak{B}_\IR$ is an orthonormal $\IC$-basis of $V$
    and $\card{\mathfrak{B}_\IR}=\card{\mathfrak{B}}$. By construction we have
    $\mathfrak{B}_\IR\subseteq V_\IR$, thus $\mathfrak{B}_\IR$ is an orthonormal $\IR$-basis of its
    $\IR$-span $U$. Since $U$ is an $\IR$-subspace of $V_\IR$ of dimension
    $\card{\mathfrak{B}_\IR}=\card{\mathfrak{B}}=\dim_\IC V=\dim_\IR V_\IR$, we have $U=V_\IR$, i.e.,
    $\mathfrak{B}_\IR$ is an orthonormal $\IR$-basis of $V_\IR$.
    \end{proof}
\end{lemma}
\begin{remark}\label{133c414f}
    The ordering \eqref{291efc26} of the basis $\mathfrak{B}$ in \thref{55efa9f6} is not uniquely
    determined. However, if $\mathfrak{B}$ is endowed with a prescribed ordering, then
    $\mathfrak{B}$ can can be uniquely reordered in the form \eqref{291efc26} by requiring
    $a_1<\cdots<a_k$ and $b_i<b_i^*$ for all $1\le i\le l$.
\end{remark}
\begin{theorem}\label{55efa9f6}
    An orthonormal $\IC$-basis of $\HS$ is explicitly given by
    \begin{equation*}
        \HSrealBasis=\left\{2^{-n/2}b_K\mid K\subseteq\IN_n\right\}\dotcup
        \left\{\frac{b_K\left(\C_{I,J}\pm\C_{J,I}\right)}{2^{(n+1-\card{I\cup J})/2}}
            \left|\begin{gathered}
                K,I,J\subset\IN_n\text{ mutually}\\
                \text{disjoint and }I<J
            \end{gathered}\right.\right\}.
    \end{equation*}
    $\HSrealBasis$ restricts to an orthonormal basis of the space $\kbOb$ of $k$-body observables
    for every $k\in\IN_0$. More specifically, an orthonormal $\IR$-basis of $\kbOb$ is given by
    \begin{equation*}
      \begin{gathered}
        \HSrealBasis_k\doteq\HSrealBasis\cap\kbOb=\left\{b_K\mid K\subseteq\IN_n\text{ and }\card{K}\le k\right\}\\
        \begin{aligned}
            \dotcup&\left\{\frac{b_K\left(\C_{I,J}\pm\C_{J,I}\right)}{2^{(n+1-\card{I\cup J})/2}}
                \left|\begin{gathered}
                    K,I,J\subset\IN_n\text{ pairwise disjoint, }I<J\\
                    \text{and }\card{I}+\card{J}+2\card{K}=2l\text{ with }0\le l\le k
                \end{gathered}\right.\right\},\\
        \end{aligned}
      \end{gathered}
    \end{equation*}
    where $I<J$ is to be understood with respect to the lexicographic ordering.
    \begin{proof}
        The first statement follows immediately from \thref{55efa9f6} applied to the
        orthonormal $\IC$-basis $\HSbasis$ as given in \thref{9faf7dc5}, which has been
        ordered according to \thref{133c414f} by defining
        \(b_K\C_{A,B}<b_L\C_{C,D}\Leftrightarrow(K,A,B)<(L,C,D)\) (lexicographic order).
    \end{proof}
\end{theorem}

\section{Alternative construction of an orthonormal basis}\label{a605ad65}

In this section, we provide an alternative construction of an orthonormal basis of $\HS$ which
restricts to an orthonormal basis of $\kbOp$ in the sense of \thref{46f8bbb2}. This construction
was already presented in \cite[Sec.~8]{Erdahl1978}, but the corresponding proofs were deferred to
a somewhat obscure reference.

Fix an orthonormal basis $\varphi_1,\ldots,\varphi_n$ of the one-particle Hilbert space
$\HilbertSpace$ and consider for $j=1,\ldots, 2n$ the operator
\begin{equation}\label{b03562f0}
    a_j\doteq \begin{cases}
        c_k^*+c_k&\text{if $j=2k$ is even},\\
        \imag\left(c_k^*-c_k\right)&\text{if $j=2k+1$ is odd}.
\end{cases}
\end{equation}
By definition, the $a_j$ are self-adjoint and, by the CAR \eqref{fee10d27}, satisfy
\begin{equation}\label{4d77cb29}
    \begin{aligned}
        \left\{a_j,a_k\right\}&=2\delta_{jk},&
        a_j^2&=\one.
    \end{aligned}
\end{equation}
Moreover, for a subset $J=\{j_1<\cdots<j_l\}\subseteq\IN_{2n}$ we define $a_J\doteq a_{j_1}\cdots
a_{j_l}$ where $a_\emptyset\doteq\one$ by convention. The following result has been suggested to us by
Gosset. We present a proof which only relies on the algebraic properties \eqref{4d77cb29} of the
elements $a_j$.
\begin{theorem}
    An orthonormal $\IC$-basis of $\HS$ is given by
    \begin{equation}
        \widetilde{\HSbasis}\doteq\left\{2^{-n/2}a_J\mid K\subseteq\IN_{2n}\right\}.
    \end{equation}
    Moreover, $\widetilde{\HSbasis}$ restricts to an orthonormal basis $\widetilde{\HSbasis_k}$ of
    $\kbOp$ for every $k\in\IN_0$, where
    \begin{equation}\label{863baee1}
        \widetilde{\HSbasis_k}\doteq\widetilde{\HSbasis}\cap\kbOp=\left\{a_J\left|
        \begin{gathered}
            J\subseteq\IN_{2n}\text{ and }\\\card{J}=2l\text{ with $0\le l\le$ k}
        \end{gathered}
        \right.\right\}.
    \end{equation}
    \begin{proof}
        We will first show that $\SP{a_J}{a_K}=2^n\delta_{JK}$ for all $J,K\subseteq\IN_{2n}$. If
        $J=K=\{j_1<\cdots<j_l\}$ then, by self-adjointness of the $a_j$ and $a_j^2=\one_\FockSpace$
        we have
        \begin{equation}
            \SP{a_J}{a_K}=\tr\{a_J^*a_J\}=\tr\{a_{j_l}\cdots a_{j_1}a_{j_1}\cdots a_{j_l}\}
            =\tr\{\one_\FockSpace\}=2^n.
        \end{equation}
        Now consider the case $J\ne K$. Without loss of generality, we may assume $J\cap K=\emptyset$ because if $i\in J\cap K$ then, by \eqref{4d77cb29},
        \begin{equation}
            \begin{aligned}
                \SP{a_J}{a_K}_\HS&=\tr\{a_J^*a_K\}
                =\pm\tr\{a_{J\setminus\{i\}}^*a_{K\setminus\{i\}}\}.
            \end{aligned}
        \end{equation}
        Moreover, by setting $I\doteq J\dot\cup K$ and noting that $\SP{a_J}{a_K}=\pm\tr\{a_I\}$, it
        suffices to show that $\tr\{a_I\}=0$ for all non-empty $I\subseteq\IN_{2n}$. First, consider
        the case where $\card{I}=l>0$ is even. Then, writing $I=\{i_1<\cdots<i_l$ we obtain, using
        \eqref{4d77cb29} and cyclicity of trace,
        \begin{equation}
            \begin{aligned}
                \tr\{a_I\}&=\tr\{a_{i_1}\cdots a_{i_l}\}=(-1)^{l-1}\tr\{a_{i_l}a_{i_1}\cdots a_{i_{l-1}}\}
                \\&=(-1)^{l-1}\tr\{a_{i_1}\cdots a_{i_l}\}=-\tr\{a_I\},
            \end{aligned}
        \end{equation}
        thus $\tr\{a_I\}=0$. On the other hand, if $\card{I}$ is odd, then consider the natural
        $\IZ_2$-grading $\FockSpace=\FockSpace_+\oplus\FockSpace_-$ on $\FockSpace$ induced by
        $\chi\doteq(-1)^{\hat{\IN}}$, i.e. $\FockSpace_\pm\doteq\ker\{\chi\mp\one\}$. By definition,
        $a_i$ is \emph{odd} with respect to this grading for any $i\in\IN_{2n}$, hence also $a_I$ is
        odd when $\card{I}$ is odd and therefore $\tr\{a_I\}=0$. We have thus proved that
        \begin{equation}
            \SP{a_J}{a_K}=2^n\delta_{JK}\qquad J,K\subseteq\IN_{2n}.
        \end{equation}
        In particular, since $\card{\widetilde{\HSbasis_k}}=2^{2n}=\dim\HS$,
        $\widetilde{\HSbasis_k}$ is an ONB of $\HS$.

        To prove \eqref{863baee1} note that, by definition, an element $a_J$ is an
        $j$-particle operator with $j\doteq\card{J}$ for any $J\subseteq\IN_{2n}$, hence $a_J$
        is a $k$-body operator if and only if $\card{J}=2l$ for some $0\le l\le k$. By
        \eqref{863baee1} and \thref{d29294a2},
        \begin{equation}
            \card{\widetilde{\HSbasis_k}}=\sum_{l=0}^k\binom{2n}{2l}=\dim\kbOp,
        \end{equation}
        thus $\widetilde{\HSbasis_k}$ is an orthonormal basis of $\kbOp$.
    \end{proof}
\end{theorem}
\begin{remark}[Relation between $\HSbasis$ and $\widetilde{\HSbasis}$]
    If $n>0$, the orthonormal bases $\widetilde{\HSbasis}$ and $\HSbasis$ are different. In fact,
    $\HSbasis\cap\widetilde{\HSbasis}=\{2^{-n/2}\one_\FockSpace\}$, since the elements of $\HSbasis$
    are homogeneous with respect to the natural grading $\FockSpace=\bigoplus_{k\ge
    0}\bigwedge^k\HilbertSpace$, whereas the elements $a_J\in\widetilde{\HSbasis}$ are inhomogeneous
    whenever $J\ne\emptyset$.
\end{remark}

\section*{Acknowledgements}
This research is supported by the German Research Foundation
(\href{http://gepris.dfg.de/gepris/projekt/399154669}{DFG Project No. 399154669}). Moreover, we
are grateful to D.~Gosset for suggesting the alternative construction presented in
Sec.~\ref{a605ad65}.

\printbibliography[heading=bibintoc]
\end{document}